\documentclass[twocolumn,showpacs,preprintnumbers,amsmath,amssymb]{revtex4}

\usepackage{graphicx}
\usepackage{dcolumn}
\usepackage{bm}

\begin{document}

\title{Synchronization on complex networks with different sorts of communities}
\author{Ming Zhao$^1$}
\email{zhaom17@mail.ustc.edu.cn}
\author{Tao Zhou$^{1,2}$}
\email{zhutou@ustc.edu}
\author{Bing-Hong Wang$^{1,3}$}
\email{bhwang@ustc.edu.cn}
\affiliation{%
$^1$Department of Modern Physics, University of Science and
Technology of China, Hefei Anhui, 230026, PR China\\
$^2$Department of Physics, University of Fribourg, Chemin du Muse
3, CH-1700 Fribourg, Switzerland \\
$^3$Shanghai Academy of System Science, Shanghai 200093, PR China
}%

\date{\today}

\begin{abstract}
In this paper, inspired by the idea that many real networks are
composed by different sorts of communities, we investigate the
synchronization property of oscillators on such networks. We
identify the communities by the intrinsic frequencies probability
density $g(\omega)$ of Kuramoto oscillators. That is to say,
communities in different sorts are functional different. For a
network containing two sorts of communities, when the community
strength is strong, only the oscillators in the same community
synchronize. With the weakening of the community strength, an
interesting phenomenon, \emph{Community Grouping}, appears: although
the global synchronization is not achieved, oscillators in the same
sort of communities will synchronize. Global synchronization will
appear with the further reducing of the community strength, and the
oscillators will rotate around the average frequency.
\end{abstract}


\maketitle

\section{Introduction}
Many social, physical and biological systems are often described by
networks. The collective dynamical behaviors of those networks are
simultaneously determined by the individual systems, the
interactions between individual systems, and the network structure.
Very recently, it is found that the structure of many large-scale
real networks are neither regular nor random, but of some common and
important characters, such as short average distance, large
clustering coefficient and the power-law degree distribution
\cite{Albert02}. Networks of such structure are called complex
networks. Besides the characters mentioned above, many real-world
networks have the so-called community structure
\cite{ComStr1,ComStr2}. A community in a network is a set of nodes,
where the edges inside are much denser than those connecting the
nodes belong to this set and the rest of the network. Examples of
community networks are numerous in biological, technical and social
systems. It is common in society that people are divided into
different groups by their sex, age, interests and so forth, and each
group can be considered as a community.

One of the most significant aims of the studies on complex networks
is to understand the effects of the network structure on the
dynamical processes, among which the synchronization of coupled
oscillators is the simplest but one of the most important ones.
Synchronization phenomenon has been observed for hundreds of years
and in a variety of field, including natural, physical, chemical and
biological systems \cite{Sync}. Because of the limitation of
knowledge on network structure, the studies of synchronization are
restricted to either on the regular lattices or on the random
networks for a long time period. Recently, with the pioneer works on
small-world \cite{SWN} and scale-free network models \cite{SFN},
much attention has been paid to the studies of synchronization on
those complex networks. Soon, scientist found delightedly that
oscillators on complex networks are much easier to synchronize than
on regular lattice with the same size and density of edges
\cite{Gade00,Wang2002a,Wang2002b,Pecora02}, which further stimulates
their enthusiasm.

Up to now, the relationship between network structure and
synchronizability
\cite{Nishikawa03,Hong04,McGraw05,ZhaoM2006,Wu06,Bernardo2,Chavez06PRE74}
is well understood and many ways to enhance the network
synchronizability
\cite{Motter05,MZKEPL,MZKAIP,Hwang05,Chavez05,ZhaoM05,ZhouC06,ZhouT06,YinCY06,ZhaoM06EPJB,Guo2007,WangXG07,Lu2007}
have been proposed (see also the review article \cite{ZhaoREV} and
the references therein). Very recently, the synchronization
properties of networks with community structure are investigated, it
is found that the connecting pattern between communities have great
effects on the synchronization process \cite{Oh05,Park06}, and the
community structure will hinder the global synchronization of
oscillators \cite{HuangPRL06,ZhouTao2007}: the stronger the
community structure the worse the global synchronizability. All the
works mentioned above concentrate on the networks with only one sort
of communities, i.e., all the communities are similar in both
function and topology. However, in real world, many networks have
communities with different sorts, an example in point is a
friendship network of children in a U.S. school
\cite{Newman03Review}. As shown in Fig. 3.4 of Ref.
\cite{Newman03Review}, there are four communities: white middle
school students, white high school students, black middle school
students, and black high school students. Clearly, the four
communities can be divided into two sorts by race: white and black,
and there are two communities in each sort. Or, those communities
can be divided by age, as middle school and high school.

In this paper, with the help of Kuramoto model
\cite{Kuramoto75,Kuramoto84,Kuramoto87,Kuramoto2,Kuramoto3}, we
investigate the synchronization properties of complex networks with
different sorts of communities. In our study, oscillators on
different sorts of communities are identified by the natural
frequencies probability density $g(\omega)$, namely, the intrinsic
frequencies of nodes on different sorts of communities are taken
differently.

This paper is organized as follows. In section
\uppercase\expandafter{\romannumeral 2}, the Kuramoto model and the
order parameter are introduced. In section
\uppercase\expandafter{\romannumeral 3}, the network model we
investigate will be described in detail. And also, we will give the
simulation results of synchronization properties of Kuramoto
oscillators on complex networks with different sorts of communities.
The conclusion remarks are drawn in section
\uppercase\expandafter{\romannumeral 4}.

\begin{figure}
\scalebox{0.75}[0.75]{\includegraphics{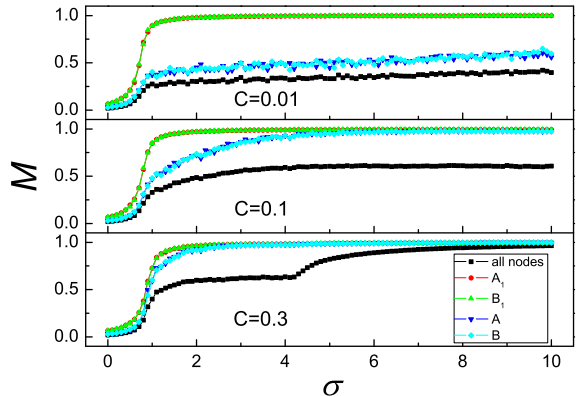}} \caption{(Color
online) The relationship between order parameter $M$ and the
coupling strength $\sigma$ for the network with two sorts of
communities, and the number of communities in those two sorts are
equal. Red circle, green triangle, blue up triangle and light blue
diamond curves represent the order parameters of A$_1$, B$_1$, A and
B, respectively. And the black square curve represents the order
parameter of the whole network. Each point is obtained from 100
independent runs.}
\end{figure}

\section{Kuramoto model and order parameter}
In this paper, we use the coupled phase oscillators, Kuramoto model
\cite{Kuramoto75,Kuramoto84,Kuramoto87,Kuramoto2,Kuramoto3}, to
analyze the collective synchronization on complex networks. A
modified Kuramoto model can be described by the coupled differential
equation \cite{ZhouTao2007}:
\begin{equation}
\frac{d\phi_i}{dt}=\omega_i-\frac{\sigma}{k_i}\sum_{j\in\Lambda_i}\sin(\phi_i-\phi_j),
\end{equation}
where $\phi_i$, $\omega_i$ are the phase and the intrinsic frequency
of node $i$, $\Lambda_i$ is $i$'s neighbor set, and $\sigma$ is the
overall coupling strength. The intrinsic frequency $\omega_i$ is
chosen from a probability density $g(\omega)$.

Usually, the intrinsic frequencies of different oscillators are
assigned differently, if there are no couplings, all the oscillators
will rotate independently, thus the phases of the oscillators are
distributed almost uniformly in the interval $[0, 2\pi]$. With the
increasing of the coupling strength, the oscillators will adhere to
each other to some extent. At last, when the coupling strength
reaches some critical point $\sigma_c$, collective synchronization
of all oscillators emerges spontaneously, although the intrinsic
frequencies of different oscillators are different.

\begin{figure}
\scalebox{0.8}[0.8]{\includegraphics{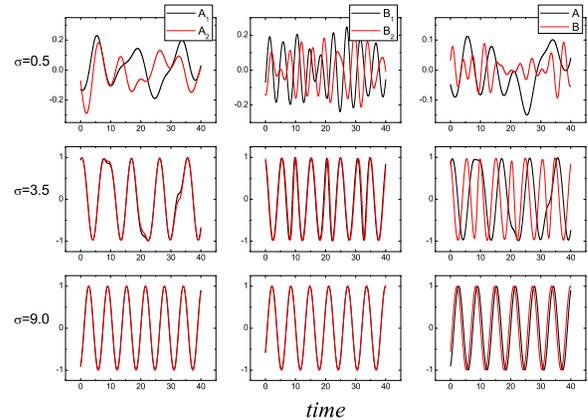}} \caption{(Color
online) The averaged sine of phases of oscillators \emph{vs.} time
for community networks with $C=0.3$ at different coupling strengths.
There are two sorts with the same number of communities in the
network.}
\end{figure}
\begin{figure}
\scalebox{0.75}[0.75]{\includegraphics{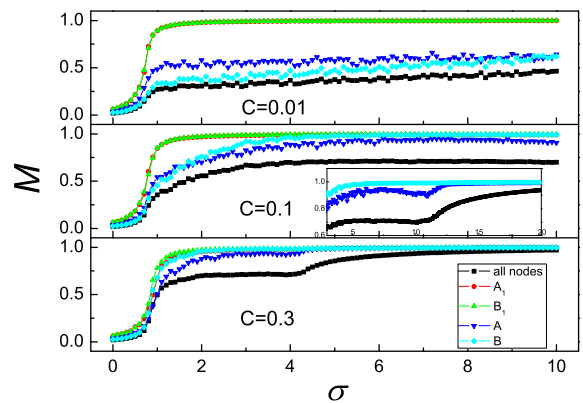}} \caption{(Color
online) The relationship between order parameter $M$ and the
coupling strength $\sigma$ for the network with two sorts of
communities. Sort A and sort B contain 3 and 7 communities,
respectively. The inset shows the change of order parameter in a
larger scale.}
\end{figure}
\begin{figure}
\scalebox{0.75}[0.75]{\includegraphics{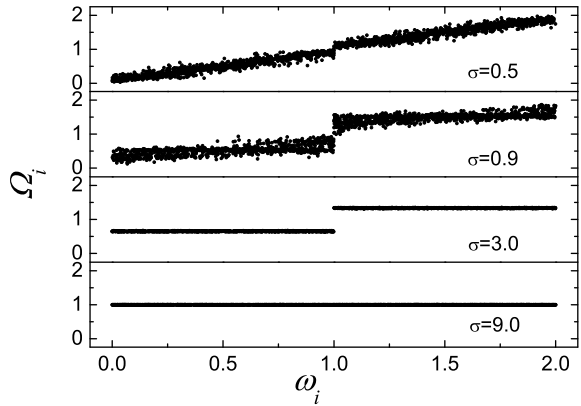}} \caption{ The
relationship between actual frequencies $\Omega_i$ and the intrinsic
frequencies $\omega_i$ for the network, which is consisted of two
sorts with equal number of communities. }
\end{figure}

To measure the synchronization phenomena, an order parameter $M$
is introduced:
\begin{equation}
M= \left[\left<\left| \frac{1}{N'}\sum_{j=1}^{N'} e^{i\phi_j}
\right|\right>\right],
\end{equation}
where $\langle\cdot\cdot\cdot \rangle$ and $[\cdot\cdot\cdot]$
denote the average over time and over different configurations,
respectively. $N'$ is the number of nodes that are taken into
account. Clearly, $M$ is of order $1/\sqrt{N'}$ if the oscillators
are completely uncoupled ($\sigma=0$), and will approach 1 if they
are all in the same phase. In our simulation, the dynamical
equations integrate using the Runge-Kutta method with step size
0.01. The order parameters are averaged over 2000 time steps,
excluding the former 2000 time steps to allow for the relaxation to
a steady state. Note that, we consider not only the order parameter
of all the oscillators in the network, but also investigate the
synchronization of partial oscillators, such as the nodes belong to
an individual community, or the nodes belong to a sort of
communities. For different cases, the sum in Eq. (2) goes over the
oscillators that are taken into account, and $N'$ is taken
accordingly.

\section{Network model and simulation results}
Our numerical simulations are based on the community network model
in Refs. \cite{ZhouTao2007,GYan07PRE}. The model starts from $n$
community cores, each core contains $m_0$ fully connected nodes.
Initially, there are no connections among different community cores.
At each time step, there are $n$ new nodes being added, each node
will attach $m$ edges to the existing nodes within the same
community core, and simultaneously $m'$ edges to the existing nodes
outside this community core. The former are internal edges, and the
latter are external edges. Similar to the evolutionary mechanism of
Barab\'{a}si-Albert networks \cite{SFN}, we assume the probability
connecting to an existing node $i$ is proportional to $i$'s degree
$k_i$. Each community core will finally become a single community of
size $N_c$, and the network size is $N = nN_c$. By using the rate
equation (similar to the analytical approaches used in Refs.
\cite{Krapivsky00,Zhou2005}), one can easily obtain the degree
distribution of the whole network, $p(k) \propto k^{-3}$. We simply
take the proportion of the external edges, $C=\frac{m'}{m+m'}$, to
measure the community strength. Clearly, a smaller $C$ corresponds
to sparser external edges thus a stronger community structure.
\begin{figure}
\scalebox{0.75}[0.75]{\includegraphics{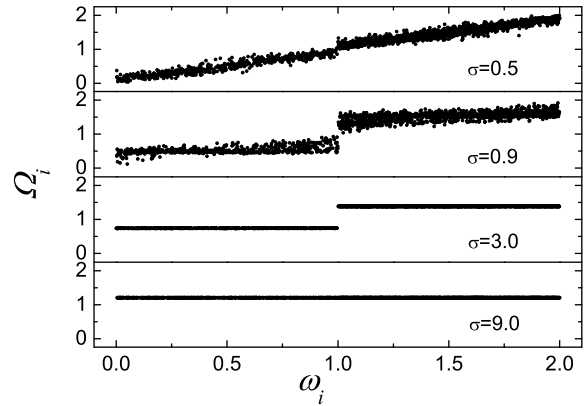}} \caption{ The
relationship between actual frequencies $\Omega_i$ and the intrinsic
frequencies $\omega_i$ for the network, which is consisted of two
sorts with different numbers of communities: sort A contains 3
communities, while sort B contains 7. }
\end{figure}

In our simulation, a network of 10 communities is taken into
account, and each community contains 200 nodes, thus there are
totally 2000 nodes in the network. The average degree is set as
$\bar{k}=6$. The communities are divided into two sorts A and B, and
each sort contains 5 communities, which are labeled as A$_1$,
$\cdots$, A$_5$ and B$_1$, $\cdots$, B$_5$, respectively. In fact,
all the ten communities have similar structure, but the intrinsic
frequencies of nodes in A are assigned randomly in $[0, 1]$, while
those in B are assigned in $[1, 2]$.

To get the synchronization property of this kind of networks, we
investigate how the order parameters in three scales, a community, a
sort of communities and the whole network, change with the coupling
strength. Figure 1 shows the simulation results. From top to bottom,
$C$ increases from 0.01, 0.1 to 0.3, corresponding to weaker and
weaker community structure. In the case $C=0.01$, with the
increasing of coupling strength, the order parameter of individual
community soon reaches 1 (the red circle and the green triangle
curves), indicating the synchronized state within one community.
However, the phases of oscillators in different communities are
different, even for those belong to the same sort (the blue up
triangle and light blue diamond lines), and the order parameter for
the whole network is much less than 1 (the black square line). The
order parameter for each community is near to 1, for community sorts
A and B, the order parameter are almost equal and much lower than
those for individual community, and the order parameter of the whole
network is the lowest one. In the case $C=0.1$, the oscillators in
the same community will also synchronize soon. With the further
increasing of coupling strength, an interesting phenomenon appears:
oscillators belong to the same sort are synchronized before the
appearance of global collective consensus. Although the communities
are equal in structure, the intrinsic frequencies divide the
communities into two groups. We call this phenomenon \emph{Community
Grouping}. In the case $C=0.3$, when the coupling strength is not
too large, the \emph{Community Grouping} can also be observed, while
for sufficiently large $\sigma$, this phenomenon disappears and the
global synchronization is achieved. In figure 2, the averaged sine
of phases of oscillators are plotted with time at steady states for
community networks with $C=0.3$. When $\sigma=0.5$, the coupling is
too weak to form any synchronization. When $\sigma=3.5$,
\emph{Community Grouping} phenomenon appears. When the coupling
strength is strong enough ($\sigma=9.0$), all the communities rotate
coherently. From figures 1 and 2 we know that only in some proper
regions of $(C,\sigma)$, the \emph{Community Grouping} phenomenon
could appear. We also investigate the network with more than two
community sorts, and find that if the numbers of communities
contained by one sort are equal, the network exhibits similar
synchronization properties to the case of two sorts.

As to networks with two sorts containing different numbers of
communities, some interesting phenomena appears. Figure 3 shows the
changing of the order parameters with the coupling strength, where
sort A and sort B contain 3 and 7 communities, respectively. For
different community strengths $C=0.01$, 0.1 and 0.3, order
parameters for individual community all reach 1 soon, showing that
oscillators within one community synchronize quickly. For the
networks with very strong community structure ($C=0.01$), even the
communities belong to the same sort could not get synchronized each
other. Meanwhile, for larger $C$, the order parameters for sort A
and B are not equal: with the coupling strength increasing from 0,
both $M_A$ and $M_B$ rise soon, but the former is faster. While,
with the further increasing of the coupling strength, the rising of
the two order parameter slow down, and the rising of $M_B$ becomes a
little faster, sooner or later $M_B$ will surpass $M_A$. This
crossing behavior of $M_A$ and $M_B$ can be observed in the panel
with $C=0.1$, and larger $C$ will make the crossing point smaller.
After the crossing point, $M_A$ will increase, decrease and increase
again, and finally reaches 1 (see the inset of Figure 3). The
abnormal change of $M_A$ can be simply explained as follows. When
$\sigma$ is small, with fewer oscillators, A will show better
coherent. For larger $\sigma$, interactions between oscillators are
strong, because of containing more community, B has stronger power
of influence, thus oscillators in A are disturbed by B. This effect
could enhance the global synchronization, but simultaneously reduce
$M_A$.

When there are couplings between oscillators, the oscillators will
not rotate at their own frequencies $\omega$ but at actual
frequencies $\Omega$, and the distribution of $\Omega$ depends on
the coupling strength and the structure of the networks. Figure 4
and 5 shows the relationships between actual frequencies and
intrinsic frequencies of each oscillator at different coupling
strength given $C=0.3$. In figure 4, there are two sorts with equal
number of communities. It can be seen that for a weaker coupling
strength, the actual frequencies $\Omega_i$ are almost equal to the
intrinsic frequencies $\omega_i$, increasing the coupling strength
$\sigma$ will make the actual frequencies of oscillators on the same
sort more consensus, till the oscillators rotate at a common
frequency near their own average frequencies, indicated by the two
parallel lines in the third panel. Further increasing the coupling
strength will make the two parallel lines nearer and nearer, and
finally, all the oscillators rotate at about the average frequency
of the whole system, indicating the achievement of the global
synchronization. Figure 5 reports the results of network consisted
of two sorts with different numbers of communities. The actual
frequencies will change alike, and the final synchronized frequency
is about the average frequency $1.2$.

\begin{figure}
\scalebox{0.28}[0.28]{\includegraphics{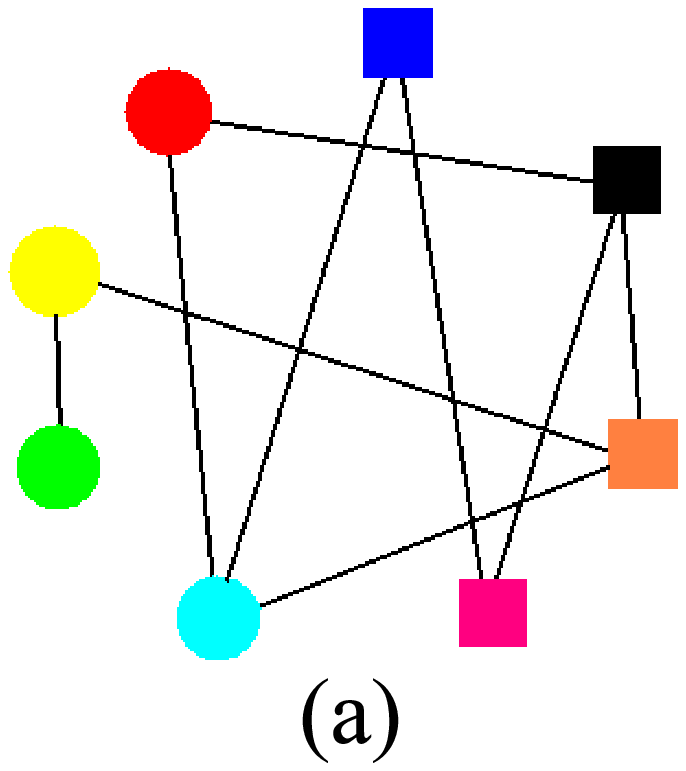}}
\scalebox{0.28}[0.28]{\includegraphics{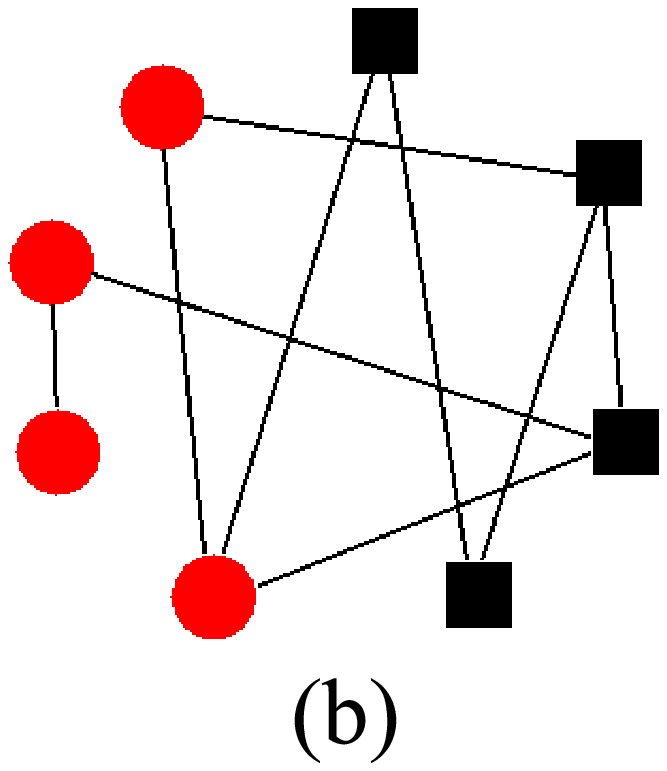}}
\scalebox{0.28}[0.28]{\includegraphics{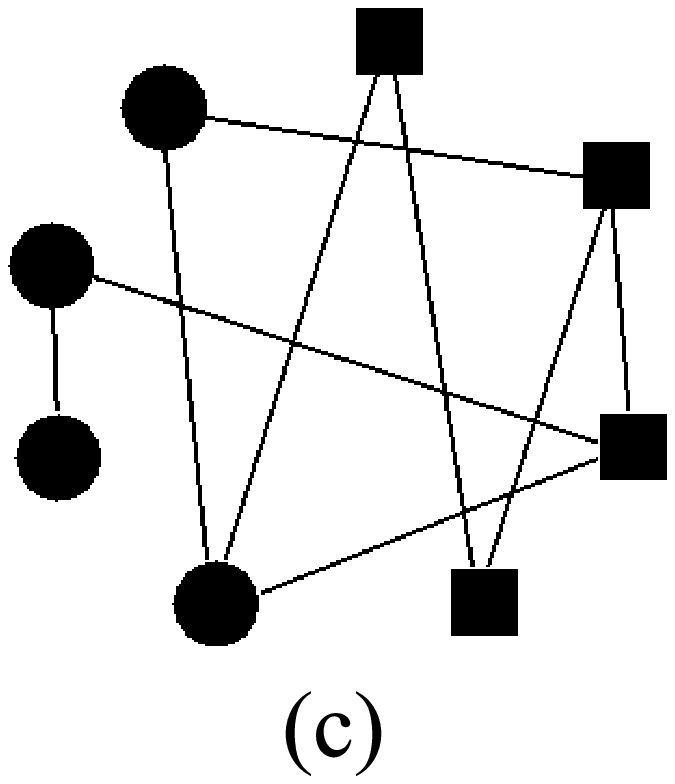}}%
\caption{(Color online) The states of the communities. The circle
and the square present two sort of community and the communities
with the same color indicate they are synchronized. (a), (b) and (c)
represent the cases of $C=0.01$, 0.1 and 0.3, as shown in the above
numerical simulations.}
\end{figure}

\section{Conclusion and Discussion}
In conclusion, we investigated the synchronization properties in
complex networks with different sorts of communities and found that
when the community structure is strong, only the oscillators in the
same community synchronize, with the weakening of the community
strength, oscillators in the same sort of communities will
synchronize independently, only when the community structure is not
evident that all the oscillators in the network can synchronize,
this is clearly shown in figure 6. When the numbers of communities
in different sorts are not equal, the sort with fewer communities
shows better consensus at weaker coupling and when the couplings
between oscillators are strong, the sort with more communities have
larger order parameter. When all the oscillators are synchronized,
they will rotate at the average frequency. The huge-size real-life
networks are generally consisted of many communities, and those
communities may be functional different. In this sense, to study the
detailed dynamical properties of network with functionally different
communities is of interesting. The current work provides a start
point on this issue, which is helpful for the in-depth understanding
of synchronization process on networks.

\begin{acknowledgments}
The authors would like to acknowledge Hui-Jie Yang and Jian-Guo Liu
for valuable discussion and suggestions, and Gang Yan for providing
the C program of network model. This work was partially supported by
the National Natural Science Foundation of China under Grant Nos.
10472116 and 10635040, the Special Research Founds for Theoretical
Physics Frontier Problems under Grant No. A0524701, and the
Specialized Program under the Presidential Funds of the Chinese
Academy of Science.
\end{acknowledgments}


\begin{thebibliography} {Albert2000}
\bibitem{Albert02} R. Albert and A. L. Barab\'{a}si, Rev. Mod. Phys. {\bf 74}, 47(2002).
\bibitem{ComStr1} M. Girvan, and M. E. J. Newnam, Proc. Natl. Acad. Sci. U.S.A. {\bf 99}, 7821(2002).
\bibitem{ComStr2} G. Palla , I. Der\'{e}nyi, I. Farkas, and T. Vicsek, Nature {\bf 435}, 814(2005).
\bibitem{Sync} S. H. Strogatz, \emph{SYNC-How the emerges from chaos in the universe, nature, and daily
life} (Hyperion, New York, 2003).
\bibitem{SWN} D. J. Watts and S. H. Strogatz, Nature {\bf 393}, 440 (1998).
\bibitem{SFN} A.-L. Barab\'{a}si and R. Albert, Science {\bf 286}, 509 (1999).
\bibitem{Gade00} P. M. Gade and C.-K. Hu, Phys. Rev. E {\bf 62}, 6409 (2000).
\bibitem{Wang2002a} X. F. Wang and G. Chen, Int. J. Bifurcation Chaos Appl. Sci. Eng. {\bf 12}, 187 (2002).
\bibitem{Wang2002b} X. F. Wang, and G. Chen, IEEE Trans. Circuits and Systems I {\bf 49}, 54 (2002).
\bibitem{Pecora02} M. Barahona and L. M. Pecora, Phys. Rev. Lett. {\bf 89}, 054101 (2002).
\bibitem{Nishikawa03} T. Nishikawa, A. E. Motter, Y.-C. Lai, and F. C. Hoppensteadt, Phys. Rev. Lett. {\bf 91}, 014101 (2003).
\bibitem{Hong04} H. Hong, B. J. Kim, M. Y. Choi, and H. Park, Phys. Rev. E {\bf 69}, 067105 (2004).
\bibitem{McGraw05} P. N. McGraw, M. Menzinger, Phys. Rev. E {\bf 72}, 015101 (2005).
\bibitem{ZhaoM2006} M. Zhao, T. Zhou, B.-H. Wang, G. Yan, H.-J. Yang, and W.-J. Bai, Physica A, {\bf 371}, 773 (2006).
\bibitem{Wu06} X. Wu, B.-H. Wang, T. Zhou, W.-X. Wang, M. Zhao, and H.-J. Yang, Chin. Phys. Lett. {\bf 23}, 1046 (2006).
\bibitem{Bernardo2} F. Sorrentino, M. di Bernardo, G. H. Cu\'ellar, and S. Boccaletti, Physica D {\bf 224}, 123 (2006).
\bibitem{Chavez06PRE74} M. Chavez, D.-U. Hwang, J. Martinerie, and S. Boccaletti, Phys. Rev. E {\bf 74}, 066107 (2006).
\bibitem{Motter05} A. E. Motter, C. Zhou, and J. Kurths, Phys. Rev. E {\bf 71}, 016116 (2005).
\bibitem{MZKEPL} A. E. Motter, C. Zhou, and J. Kurths, Europhys. Lett. {\bf 69}, 334 (2005).
\bibitem{MZKAIP} A. E. Motter, C. Zhou, and J. Kurths, AIP Conf. Proc. {\bf 776}, 201 (2005).
\bibitem{Hwang05} D.-U. Hwang, M. Chavez, A. Amann, and S. Boccaletti, Phys. Rev. Lett. {\bf 94}, 138701 (2005).
\bibitem{Chavez05} M. Chavez, D. -U. Hwang, A. Amann, H. G. E. Hentschel, and S. Boccaletti, Phys. Rev. Lett. {\bf 94}, 218701 (2005).
\bibitem{ZhaoM05} M. Zhao, T. Zhou, B.-H. Wang, and W.-X. Wang, Phys. Rev. E {\bf 72}, 057102 (2005).
\bibitem{ZhouC06} C. Zhou and J. Kurths, Phys. Rev. Lett. {\bf 96}, 164102 (2006).
\bibitem{ZhouT06} T. Zhou, M. Zhao, and B.-H. Wang, Phys. Rev. E {\bf 73}, 037101 (2006).
\bibitem{YinCY06} C.-Y. Yin, W.-X. Wang, G. Chen, and B.-H. Wang, Phys. Rev. E {\bf 74}, 047102 (2006).
\bibitem{ZhaoM06EPJB} M. Zhao, T. Zhou, B.-H. Wang, Q. Ou, and J. Ren, Eru. Phys. J. B {\bf 53}, 375 (2006).
\bibitem{Guo2007} Q. Guo, J.-G. Liu, R.-L. Wang, X.-W. Chen, and Y.-H. Yao, Chin. Phys. Lett. {\bf 24}, 2437 (2007).
\bibitem{WangXG07} X. Wang, Y.-C. Lai, and C. H. Lai, Phys. Rev. E {\bf 75}, 056205 (2007).
\bibitem{Lu2007} Y.-F. Lu, M. Zhao, T. Zhou, and B.-H. Wang, arXiv: 0708.0863.
\bibitem{ZhaoREV} M. Zhao, T. Zhou, G. Chen, and B.-H. Wang, Front. Phys. China {\bf 2}, 460 (2007).
\bibitem{Oh05} E. Oh, K. Rho, H. Hong, B. Kahng, Phys. Rev. E {\bf 72}, 047101 (2005).
\bibitem{Park06} K. Park, Y.-C. Lai, S. Gupte, J.-W. Kim, Chaos {\bf 16}, 015105 (2006).
\bibitem{HuangPRL06} L. Huang, K. Park, Y.-C. Lai, L. Yang, and K. Yang, Phys. Rev. Lett. {\bf 97}, 164101 (2006).
\bibitem{ZhouTao2007} T. Zhou, M. Zhao, G. Chen, G. Yan, and B. -H. Wang, Phys. Lett. A {\bf 368}, 431 (2007).
\bibitem{Newman03Review} M. E. J. Newman, SIAM Review, {\bf 45}, 167 (2003).
\bibitem{Kuramoto75} Y. Kuramoto, in \emph{Internaltional Symposium on Mathematical Problems in Theoretical
Physics}, edited by H. Araki, Lecture notes in Physics No. 30
(Springer, New York, 1975).
\bibitem{Kuramoto84} Y. Kuramoto, \emph{Chemical Oscillations, Wave and Turbulence} (Springer-Verlag, Berlin, 1984).
\bibitem{Kuramoto87} Y. Kuramoto, and I. Nishikawa, J. Stat. Phys. {\bf 49}, 569 (1987).
\bibitem{Kuramoto2} A. Pikovsky, \emph{Synchronization} (Cambridge University Press, Cambridge, 2001).
\bibitem{Kuramoto3} J. A. Acebr\'on, L. L. Bonilla, C. J. P. Vicente, F. Ritort, and R. Spigler, Rev. Mod. Phys. {\bf 77}, 137 (2005).
\bibitem{GYan07PRE} G. Yan, Z. -Q. Fu, J. Ren, and W. -X. Wang, Phys Rev E {\bf 75}, 016108 (2007).
\bibitem{Krapivsky00} P. L. Krapivsky, S. Render, and F. Leyvraz, Phys. Rev. Lett. {\bf 85}, 4629(2000).
\bibitem{Zhou2005} T. Zhou, G. Yan, and B. -H. Wang, Phys. Rev. E
{\bf 71}, 046141 (2005).
\end{thebibliography}
\end{document}